\documentstyle[11pt,newpasp,twoside,epsf]{article}
\begin{document}
\title{Relativistic Jets from Collapsars: Gamma-Ray Bursts}
 \author{Weiqun Zhang and S. E. Woosley}
\affil{Department of Astronomy and Astrophysics, University of
 California, Santa Cruz, CA 95064, USA}

\begin{abstract}
Growing observational evidence supports the proposition that gamma-ray
bursts (GRBs) are powered by relativistic jets from massive helium
stars whose cores have collapsed to black holes and an accretion disk
(collapsars).  We model the propagation of relativistic jets through
the stellar progenitor and its wind using a two-dimensional special
relativistic hydrodynamics code based on the PPM formalism.  The jet
emerges from the star with a plug in front and a cocoon surrounding
it.  During its propagation outside the star, the jet gains high
Lorentz factor as its internal energy is converted into kinetic energy
while the cocoon expands both outwards and sideways.  External shocks
between the cocoon and the stellar wind can produce $\gamma$-ray and
hard x-ray transients.  The interaction of the jet beam and the plug
will also affect both of them substantially, and may lead to
short-hard GRBs. Internal shocks in the jet itself may make long-soft
GRBs.
\end{abstract}

\section{Introduction}

Although there is no universally agreed upon central engine powering
gamma-ray bursts (GRBs), growing evidence supports the association of
least the long-soft GRBs (all those whose counterparts have been
localized) with the death of massive stars.  This evidence includes
the association of GRBs with regions of massive star formation (Bloom,
Kulkarni, \& Djorgovski 2002) and ``bumps'' in the optical afterglows
of several GRBs that have been related to the light curves of Type I
supernovae (e.g., Bloom et al. 2002; Garnavich et al. 2002).  In
addition, GRB 980425 has been associated with SN 1998bw (e.g., Iwamoto
et al. 1998; Woosley, Eastman, \& Schmidt 1999).  Frail et al. (2001),
Panaitescu \& Kumar (2001) have studied beaming angles and energies of
a number of GRBs and have found that the central engines of GRBs release
supernova-like energies.  Given these discoveries, the currently favored
models are all based upon the collapse of massive stars and their
byproducts, one of which is a relativistic jet.

Among those models involving massive stars, the collapsar model
(Woosley 1993; MacFadyen \& Woosley 1999) has become a favorite.  A
``collapsar'' is a rotating massive star whose iron core has collapsed
and formed a black hole and an accretion disk.  In this model, the
central black hole and the disk use neutrinos or magnetic fields to
extract part of the gravitational potential or rotational energy and
form powerful relativistic jets along the polar axes.  Jets from
collapsars have been studied numerically in both Newtonian (MacFadyen
\& Woosley 1999; MacFadyen, Woosley, \& Heger 2001) and relativistic
simulations (Aloy et al. 2000; Zhang, Woosley, \& MacFadyen 2002) and
it has been shown that the collapsar model is able to explain many of
the observed characteristics of GRBs.  These previous studies have
also raised issues which require further examination, especially with
higher resolution.  For instance, the emergence of the jet and its
interaction with the material at the stellar surface and the stellar
wind will definitely lead to some sort of ``precursor'' activity.  The
long term dynamics of the jet is critical in producing the observed
gamma-rays and afterglows.  We have recently carried out
multi-dimensional calculations to address some of these issues.

\section{Numerical Methods and Initial Conditions}

A collapsar is formed when the core of a massive star collapses to a
black hole and an accretion disk. The interaction of this disk with
the hole, through processes that are still poorly understood, produces
jets with a high energy to mass ratio. For our simulations, we are
concerned primarily with the propagation of these relativistic jets,
their interactions with the star and the stellar wind, and the
observational implications, and not so much with how they are born.
We model the propagation of relativistic jets inside and just outside
collapsars using a multi-dimensional relativistic hydrodynamics code
that has been used previously to study relativistic jets in the
collapsar environment (Zhang et al. 2002).  Briefly, our code employs
an explicit Eulerian Godunov-type method developed by Aloy et
al. (1999).  Three-dimensional numerical simulations of relativistic
jets in collapsars are still rather expensive, so the present study
consists of a series of two-dimensional calculations with high
resolution.

Our initial model is a $15\,M_{\sun}$ helium star calculated by Heger
\& Woosley (2002).  The helium star has been evolved to iron core
collapse.  This presupernova model is then remapped into our
two-dimensional cylindrical grid $(r, z)$, which consists of 1500
zones in $0\le r \le 6 \times 10^{11}\,{\mathrm cm}$ and 2275 zones in
$10^{10}\,{\mathrm cm} \le z \le 2 \times 10^{12}\,{\mathrm cm}$.  In
one of the models, we used a larger grid in $r$-direction, 2375 zones
for $0\le r \le 2 \times 10^{12}\,{\mathrm cm}$.  In each case, the
zoning is nonuniform with higher resolution in the inner region.  The
radius of the initial helium star is about $8 \times 10^{10}\,{\mathrm
cm}$ and the surface of the star is very finely zoned.  Outside the
star, the background density, which comes from the stellar wind, is
assumed to be $\sim R^{-2}$, and the density at $R =
10^{11}\,{\mathrm cm}$ is set to $5 \times 10^{-11}\,{\mathrm
g}\,{\mathrm cm}^{-3}$, here $R$ is the distance to the center of the
star.  This corresponds to a mass loss rate of $\sim
10^{-5}\,M_{\sun}\,{\mathrm yr}^{-1}$ and a velocity of $\sim
1000\,{\mathrm km}\,{\mathrm s}^{-1}$ at $10^{11}\,{\mathrm cm}$.

\section{Results}

We presume that a highly relativistic jet has already formed inside
the inner boundary of our computational grid and study its evolution
after it enters the grid.  In particular, an axisymmetric jet is
injected along the rotation axis through the inner boundary within a
radius, $r_0$, here taken to be a free parameter.  The initial jet is
additionally defined by its total energy (excluding rest mass energy)
flux per jet, $\dot{E}$, initial Lorentz factor, $\Gamma_0$, and the
ratio of its kinetic energy to total energy, $f_0$.  We have run a
series of calculations: (A) $r_0 = 9 \times 10^{8}\,{\mathrm cm}$,
$\dot{E} = 3 \times 10^{50}\,{\mathrm erg}\,{\mathrm s}^{-1}$,
$\Gamma_0 = 5$, $f_0 = 0.025$; (B) $r_0 = 9 \times 10^{8}\,{\mathrm
cm}$, $\dot{E} = 10^{50}\,{\mathrm erg}\,{\mathrm s}^{-1}$, $\Gamma_0
= 10$, $f_0 = 0.05$; and (C) $r_0 = 9 \times 10^{8}\,{\mathrm cm}$,
$\dot{E} = 5 \times 10^{49}\,{\mathrm erg}\,{\mathrm s}^{-1}$,
$\Gamma_0 = 5$, $f_0 = 0.025$.  There presently exists no rigorous
calculation of the jet formation process.  However, a total energy of
order $10^{51} - 10^{52}$ ergs is reasonable.  The typical duration of a
long GRB is $\sim 20$ seconds.  In our calculations, the jet is left
on for 20 seconds, and then gradually shut off (linearly with time).
We know from observations that the GRB outflows are narrowly beamed.
In our calculations, the radius of the jet is chosen to have a
half-opening angle of $\sim 5{\deg}$ at $10^{10}\,{\mathrm cm}$.  It
is more difficult to estimate the Lorentz factor and the ratio of
kinetic energy to total energy.  As our previous studies showed, a
jet that starts with a high Lorentz factor $\sim 50$ at $2000\,{\mathrm
km}$ will be shocked and its Lorentz factor will become $\sim 10$ at
$10^{10}\,{\mathrm cm}$ (Zhang et al. 2002).  For the parameters
chosen, the jet starts with a Lorentz factor of 5 or 10 at
$10^{10}\,{\mathrm cm}$, and would have a final Lorentz factor of $\sim
180$ if all internal energy is converted into kinetic energy.

\subsection{Jets Inside Stars and the Emergence of Jets }

In all models, the jet propagates along the rotational axis.  After a
short time, the jet consists of a supersonic jet beam; a cocoon made
of shocked jet material and shocked medium; a terminal bow shock; a
contact discontinuity; and backflows.  A snapshot of Model B is shown
in Figure~1. In agreement with our previous studies (Zhang et
al. 2002), the jet in both models is narrowly collimated and its core,
very thin.  Interaction of the jet with the star and the cocoon
imprints considerable time structure on the flow.  As the jet passes
through the star, it also explodes it.

\begin{figure}
\plotone{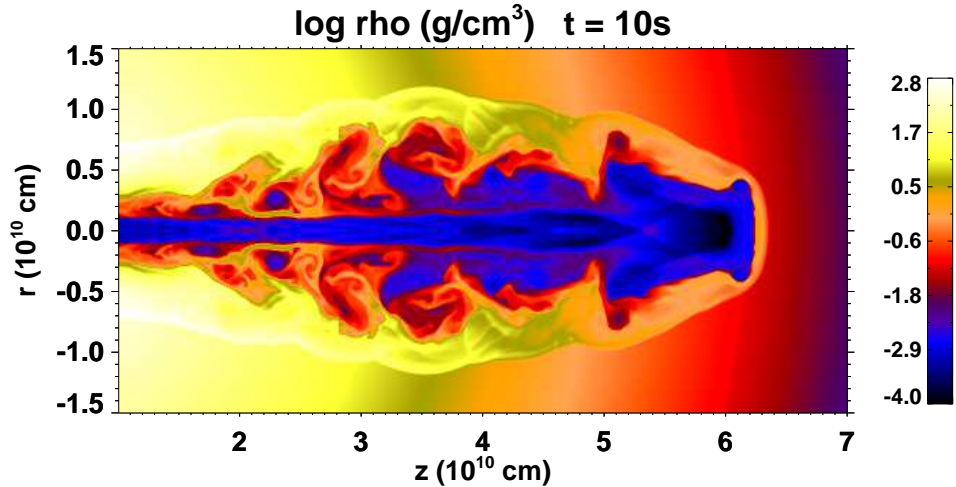}
\caption{Density structure in the local rest frame for Model B at
  10\,s.}
\end{figure} 

Eventually, the jet breaks out the star. The average velocities of the
head of the jet just before breakout are $9 \times 10^{9}$, $7 \times
10^{9}$, and $5 \times 10^{9}\,{\mathrm cm}\,{\mathrm s}^{-1}$, for
Models A, B, and C, respectively.  As expected, the jet accelerates as
it moves outwards because the density and pressure of the star
decrease quickly, especially at the surface of the star.  As the jet
passes through the star, an over-pressurized cocoon is formed.  This
cocoon escapes from the star along with the jet beam and accelerates
as its internal energy is converted into kinetic energy by adiabatic
expansion (Figure~2; see also Ramirez-Ruiz, Celotti, \& Rees 2002).
After the breakout into the low density stellar wind, the cocoon is no
longer confined.  The evolution of the jet and its cocoon in the
stellar wind will be further discussed in \S~3.2.

\begin{figure}
\plotone{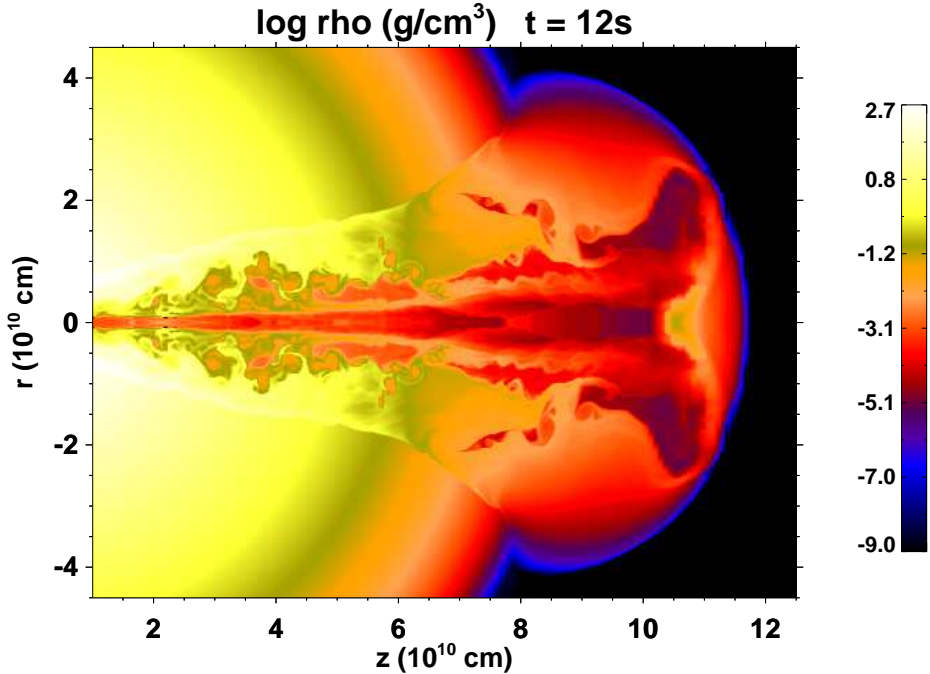}
\caption{Density structure in the local rest frame for Model B at
  12\,s. The jet is breaking out the star. At the head of the jet, the
  cocoon is expanding sideways very quickly.}
\end{figure} 

The velocity of the head of the jet in the star is subrelativistic,
and the jet beam is highly relativistic.  Shocked medium and shocked
jet material piles up near the contact discontinuity and feeds the
cocoon.  This forms a ``plug'' at the head of the jet.  After
breakout, the initially subrelativistic plug will be moving at a
Lorentz factor of $\sim 10-20$ since it is accelerated by the highly
relativistic jet and the stellar wind cannot hinder its movement.  The
interaction of the ``plug'' and the jet and its implications for GRBs
will be discussed in \S~3.2 and \S~4 (see also Waxman \&
M\'{e}sz\'{a}ros 2002).

\subsection{Jets in the Stellar Wind}

After it breaks out the star, both the jet and its cocoon are loaded
with a lot of internal energy.  Although its current Lorentz factor is
only about 10, the final Lorentz factor of the jet can be $\sim 200$
if it can expand adiabatically to gain its terminal Lorentz factor.
The Lorentz factor and density at the end of our simulation for Model
A is shown in Figure~3.  In all models, the average Lorentz factor in
the jets becomes more than 100 at the end of the simulations (t=
70\,s), and there are still some internal energy remaining in the jets
at that moment.  When they are viewed on the axis, the jets have an
isotropic equivalent total energy of $3 \times 10^{54}$, $4 \times
10^{54}$, and $5 \times 10^{53}\,{\mathrm erg}$, for Models A, B, and
C, respectively\footnote{Note that the jets are beamed into $\sim
5{\deg}$.  The isotropic equivalent energy is derived by assuming an
isotropic fireball instead of a jet}.  This energy should be enough to
power a normal GRB even if the efficiency for making $\gamma$-rays is
very small.

\begin{figure}
\plotone{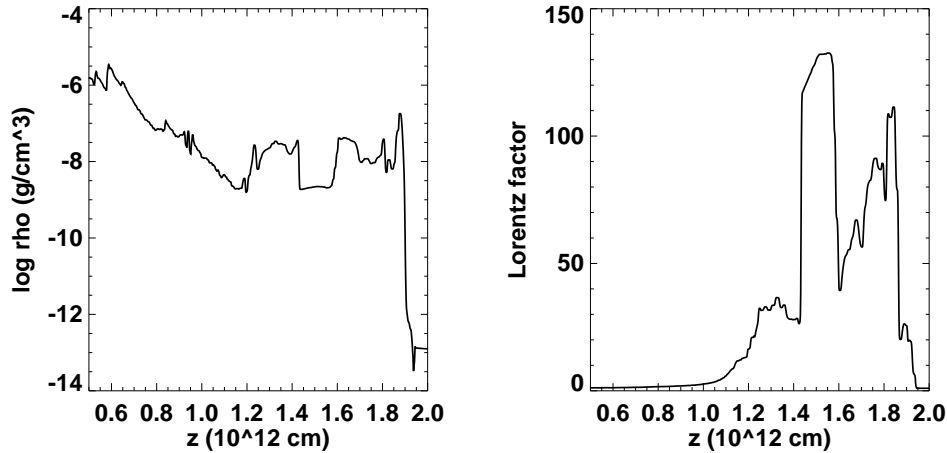}
\caption{Density and Lorentz factor along the jet axis for Model A at
  70\,s.  The Lorentz factor of the jet contains significant
  variabilities, which are very important for making $\gamma$-rays via
  internal shocks.  Note the ``plug'' at the head of the jet.  The
  plug has a moderate Lorentz factor.}
\end{figure} 

After the jet breaks out the star, its cocoon expands both outwards
and sideways (Figure~2). An interesting effect of the cocoon is that
it prevents the jet beam from expanding sideways.  Without the cocoon,
a bare ``hot'' jet with a Lorentz of about 10 will inevitably expand
sideways (Zhang et al. 2002).  The cocoon emerges out the star with an
energy close to the total energy injected at the base of the jet
before the jet emerges from the star (Ramirez-Ruiz et al. 2002).  The
conversion of its internal energy into kinetic will give the cocoon a
Lorentz factor of $\sim 5 - 10$ and the cocoon can reach up to
$30{\deg}$.  For instance, the isotropic equivalent energy of
relativistic material at $30{\deg}$ is about $10^{50}\,{\mathrm erg}$
for Model C. In Models A and B it is an order of magnitude greater.

The jet breaks out the star with a plug in front of it.  The plug is
made of the shocked stellar and jet material.  After breakout, the
plug has comparable density to that of the jet, and a Lorentz factor
of about 10-20 (see e.g., Figure~3).  In Model A, the total rest mass
in the plug at $70\,{\mathrm s}$, is about $10^{28}\,{\mathrm g}$,
which is $\sim 1/5$ of that of the jet.  They are $6 \times 10^{27}$
and $2 \times 10^{27}\,{\mathrm g}$, about half of that of the jet,
for Models B and C, respectively.  These massive plugs will surely
alter the structure of the GRB jets.  The interaction of the jet and
the plug will accelerate the plug. Meanwhile, the jet will be slowed
down.  Part of its kinetic energy will be converted into its internal
energy.  This interaction may have important implications for
observations.

\section{Discussion}

Our special relativistic calculations show that a jet originating near
the center of a collapsing massive will emerge while still carrying
almost all the energy injected at the center.  After it breaks out of the
star, the jet will move forward almost freely, no more cocoon will be
generated.  The standard fireball model requires not only high Lorentz
factor, but also the Lorentz factor to vary rapidly.  Our simulations
show promise in satisfying these constraints.  However, due to the low
resolution ($\Delta z = 1.6 \times 10^{9}\,{\mathrm cm}$) outside the
star, it is hard to resolve those small scale variabilities and the
numerical viscosity gradually smooths the time structure in the
jet.

We have found that the cocoon can expand sideways up to $\sim30{\deg}$
after it emerges from the star.  The deceleration of the cocoon by
external shocks will produce $\gamma$-rays and x-rays (see also
Ramirez-Ruiz et al. 2002).  The ``anomalous'' GRB 980425 might be a
cocoon viewed from a large angle.  It might also contribute to the
afterglows and precursors of a normal GRB and may be the origin of
recently discovered hard x-ray flashes (Heise et al. 2001; Kippen et
al 2002). These flashes have many properties of GRBs (energy,
isotropy, approximate duration, distribution with redshift), but have
a much softer spectrum (and so far no optical afterglow). The
association of this kind of softer transient with gamma-ray bursts
seen off-axis has been predicted by our group for many years (Woosley
et al 1999; Woosley \& MacFadyen 1999; Woosley 2000, 2001).

Behind the bow shock, there is a plug in front of the jet
beam.  After breakout, the main jet beam continues to push and
accelerate this plug. Meanwhile, the jet beam is slowed down as its
energy is transfered to the plug and part of its kinetic energy is
converted into its internal energy.  The basic picture of a fast
moving jet beam pushing a ``slowly'' moving plug is that there is a
reverse shock, which moves backward and slows down the jet beam, a
contact discontinuity, and a forward shock.  Since the plug has a
limited length, the forward shock will reflect at the end of the plug.
This further complicates the dynamics (e.g., Waxman \&
M\'{e}sz\'{a}ros 2002).

At the end of our simulation, the jet and the plug are moving
relativistically and the stellar wind has not been able to decelerate
the jet, so we can make some analytic estimates treating the jet
and the plug as a spherical symmetric fireball.

Where will the jet and the plug become optically thin?  In our case,
the opacity in the jet and the plug is dominated by Thomson
scattering.  The optical depth is $\tau = \kappa\,\Sigma =
\kappa\,\rho\,\Delta r = \kappa\,(M / 4 \pi r^2) = \tau_0\,(r_0/r)^2$,
here, $\kappa \sim 0.2\,{\mathrm cm}^2\,{\mathrm g}^{-1}$, and
$\tau_0$ is the initial optical depth at $r_0$.  From our results at
$t = 70\,{\mathrm s}$, the plug will become optically thin at
$r_{\mathrm th,p} = 2.0\times 10^{14}$, $2.6 \times 10^{14}$, and $1.7
\times 10^{14}\,{\mathrm cm}$, for Models A, B, and C,
respectively. And the jet will become optically thin at $r_{\mathrm
th,j} = 5.8\times 10^{14}$, $7.6 \times 10^{14}$, and $2.9 \times
10^{14}\,{\mathrm cm}$, for Models A, B, and C, respectively.  When it
becomes optically thin, the interaction of the plug with the jet, will
produce hard emission.  In fact, it is possible that this emission
could be a short-hard GRB (as defined in Fishman \& Meegan 1995).

Where will the jet catch up the plug and be decelerated by the plug?
This is a very critical question. During the catch-up period, some of
the kinetic energy of the jet is converted into internal energy.  If the
deceleration happens in the optically thick regime, the increased
internal energy can still be converted back into kinetic energy.  If,
however, it happens in the optically thin regime, the increased
internal energy will become radiation energy via this special kind of
``internal shock'' and escape from the jet.  There may not be enough
kinetic energy left and high enough Lorentz factor to make
$\gamma$-rays via ``normal'' internal shocks.  In this case, the short
hard precursors from the plug may dominate at X-ray and $\gamma$-ray
wavelengths, and a short hard GRB is likely to be seen.  Numerical
simulations on the catch-up process are currently underway.  We
analytically estimate the radius, $r_{\mathrm cat}$ , at which the
catch-up process will happen.  For simplicity we assume that the jet
moves at a Lorentz factor of 100, the plug moves at a Lorentz factor
of 20, and the length of the jet is about $5 \times 10^{11}\,{\mathrm
cm}$, then we get $r_{\mathrm cat} \sim 5 \times 10^{14}\,{\mathrm
cm}$.  This radius is comparable to the radius where the jet becomes
optically thin.  Our simulations show a tendency for the energy in the
plug relative to that in the jet behind the plug to be larger when the
total energy is less (see \S~3.2).  This suggests that less energetic
jets may be more likely to make short-hard GRBs. The near coincidence
of the masses of the plug and jet and of the radii where the two share
their energy with the gamma-ray photosphere suggests that there may be
cases where the most prominent display comes from the plug and others
where it still comes from internal shocks in the jet (or at the
jet-plug interface).

Where will the jet be decelerated by the stellar wind?  The
deceleration by the stellar wind happens when the jet sweeps up
$1/\Gamma$ of its rest mass, here $\Gamma$ is the Lorentz factor of
the jet.  Assuming a stellar wind that has a mass loss rate of $\sim
10^{-5}\,M_{\sun}\,{\mathrm yr}^{-1}$ and a velocity of $\sim
1000\,{\mathrm km}\,{\mathrm s}^{-1}$ at $10^{11}\,{\mathrm cm}$, the
deceleration radius, $r_{\mathrm dec}$, is $3 \times 10^{16}\,{\mathrm cm}$ for
Model A.  And they are $4 \times 10^{16}$ and $8 \times
10^{15}\,{\mathrm cm}$, for Models B and C, respectively.  The
deceleration by external shocks with the stellar wind always happens
after the above events. This justifies our one-dimensional analytic
calculations because the effects of the sideways expansion are
negligible when the jets are highly relativistic.

In our numerical simulations, We have followed the propagation of jets
to $r=2 \times 10^{12}\,{\mathrm cm}$.  However, many interesting
events which are directly related to observations happen at large
radius. So the long time evolution of the jets in the stellar wind is
very critical.  It will help us make many testable predictions.
Numerical calculations which follow the long time evolution of the
jets are under way.  Hopefully, they will further strengthen our
knowledge on relativistic outflows of GRBs.

It is also very important to repeat our simulations in
three-dimensional Cartesian coordinates.

\acknowledgments

This research has been supported by NASA (NAG5-8128, NAG5-12036, and
MIT-292701) and the DOE Program for Scientific Discovery through
Advanced Computing (SciDAC; DE-FC02-01ER41176). We are indebted to
Alex Heger for providing the precollapse model used in this
calculation and to illuminating conversations with Chris Matzner,
Martin Rees, and Enrico Ramirez-Ruiz.


\begin{references}

\reference 
Aloy, M. A., Ib\'{a}\~{n}ez, J. M$^{\underline{\mbox{a}}}$., Mart\'{\i},
J. M$^{\underline{\mbox{a}}}$., \& M\"{u}ller, E. 1999, \apjs, 122, 151

\reference
Aloy, M. A., M\"{u}ller, E., Ib\'{a}\~{n}ez,
J. M$^{\underline{\mbox{a}}}$., Mart\'{\i},
J. M$^{\underline{\mbox{a}}}$., \& MacFadyen, A. I. 2000, \apjl, 531,
L119

\reference
Bloom, J. S., Kulkarni, S. R., Djorgovski, S. G., Eichelberger, A. C.,
Cote, P., Blakeslee, J. P., Odewahn, S. C., Harrison, F. A., et
al. 1999, Nature, 401, 453

\reference
Bloom, J. S., Kulkarni, S. R., Price, P. A., Reichart, D., Galama,
T. J., Schmidt, B. P., Frail, D. A., Berger, E., et al. 2002, \apjl,
572, L45

\reference
Bloom, J. S., Kulkarni, S. R., \& Djorgovski, G. 2002, \aj,
123, 1111

\reference
Fishman, G. J., \& Meegan, C. A. 1995, ARA\&A, 33, 415.

\reference
Frail, D., et al. 2001, \apjl, 562, L55

\reference
Galama, T. J., Tanvir, N., Vreeswijk, P. M., Wijers, R. A. M. J.,
Groot, P. J., Rol, E., van Paradijs, J., Kouveliotou, C.  et al. 2000,
ApJ, 536, 185

\reference
Garnavich, P. M., et al. 2002, \apj, submitted, astro-ph/0204234

\reference 
Heger, A., \& Woosley, S. E. 2002, to appear in Proc. Woods Hole GRB
meeting, ed. R. Vanderspek, astro-ph/0206005

\reference
Heise, J., in't Zand, J., Kippen, R. M., Woods, P. M. 2001, GRBs in
the Afterglow Era, eds. Costa, Frontera, \& Hjorh, ESO Astrophysics
Symposia, (Springer), 16,  astro-ph/0111246.

\reference
Iwamoto, K., et al. 1998, Nature, 395, 672

\reference
Kippen, R. M., Woods, P. M, Heise, J., in 't Zand, J. J. M., Briggs,
M. S., \& Preece, R. D. 2002, in proceedings of the Woods Hole GRB
Workshop, ed. R. Van der Speck, in press, astro-ph/0203114.

\reference
MacFadyen, A. I., \& Woosley, S. E. 1999, \apj, 524, 262

\reference
MacFadyen, A. I., Woosley, S. E., \& Heger, A. 2001, \apj, 550, 410

\reference
Panaitescu, A., \& Kumar, P. 2001, \apjl, 560, L49 

\reference
Ramirez-Ruiz, E., Celotti, A., \& Rees, M. J. 2002, MNRAS, in press 
astro-ph/0205108

\reference
Waxman, E., \& M\'{e}sz\'{a}ros, P. 2002, \apj, submitted, astro-ph/0206392

\reference
Woosley, S. E. 1993, \apjl, 405, L273

\reference
Woosley, S. E., Eastman, R. G., \& Schmidt, B. 1999, \apj, 516, 788

\reference
Woosley, S. E. \& MacFadyen, A. I. 1999, A\&AS, 138, 499 

\reference
Woosley, S. E. 2000, GRBs, 5th Huntsville Symposium, eds. Kippen,
Mallozzi, \& Fishman, AIP, Vol 526, 555

\reference
Woosley, S. E. 2001, GRBs in the Afterglow Era, eds. Costa, Frontera, 
\& Hjorh, ESO Astrophysics Symposia, (Springer), 555

\reference
Zhang, W., Woosley, S. E., \& MacFadyen, A. I. 2002, \apj, in press,
astro-ph/0207436

\end{references}
\end{document}